\documentclass[12pt,preprint]{aastex62}
\usepackage{natbib}
\usepackage{xspace}

\newcommand{\be}{\begin{equation}}
\newcommand{\ee}{\end{equation}}

\newcommand{\Fermi}{\textit{Fermi}\xspace}
\newcommand{\Fermilat}{\textit{Fermi}-LAT\xspace}

\shorttitle{4FGL DR4}
\shortauthors{\Fermilat collaboration}

\begin{document}

 

\title{\Fermi Large Area Telescope Fourth Source Catalog Data Release 4 (4FGL-DR4)}
    

\author{J.~Ballet}
\email{jean.ballet@cea.fr}
\affiliation{Universit\'e Paris Saclay and Universit\'e Paris Cit\'e, CEA, CNRS, AIM, F-91191 Gif-sur-Yvette, France}
\author{P.~Bruel}
\email{Philippe.Bruel@llr.in2p3.fr}
\affiliation{Laboratoire Leprince-Ringuet, \'Ecole polytechnique, CNRS/IN2P3, F-91128 Palaiseau, France}
\author{T.~H.~Burnett}
\email{tburnett@u.washington.edu}
\affiliation{Department of Physics, University of Washington, Seattle, WA 98195-1560, USA}
\author{B.~Lott}
\email{lott@cenbg.in2p3.fr}
\affiliation{Universit\'e Bordeaux, CNRS, LP2I Bordeaux, UMR 5797, F-33170 Gradignan, France}
\collaboration{The \Fermilat collaboration}

\begin{abstract}
We present an incremental version (4FGL-DR4, for Data Release 4) of the fourth \Fermilat catalog containing 7194 $\gamma$-ray sources.
Based on the first 14 years of science data in the energy range from 50~MeV to 1~TeV, it uses the same analysis methods as the 4FGL-DR3 catalog did for 12 years of data, with only a few improvements. The spectral parameters, spectral energy distributions, light curves and associations are updated for all sources.

We add four new extended sources and modify two existing ones. Among the 6658 4FGL-DR3 sources, we delete 14 and change the localization of 10, while 32 are newly associated, eleven associations are changed and three associations are discarded. We add 546 point sources, among which 8 are considered identified and 229 have a plausible counterpart at other wavelengths.
Most are just above the detection threshold, and 14 are transient sources below the detection threshold that can affect the light curves of nearby sources.
\end{abstract}


\section{Introduction}
\label{introduction}

The \Fermi Large Area Telescope (LAT) has been surveying the high-energy $\gamma$-ray sky since 2008 \citep{LAT09_instrument} and the LAT Collaboration has published a succession of source catalogs based on comprehensive analyses of LAT data.
The fourth source catalog \citep[4FGL,][]{LAT20_4FGL} was derived from analysis of the first 8 years of LAT science data. It was followed by two incremental versions (Data Releases) called DR2 \citep{LAT20_4FGLDR2} and DR3 \citep{LAT22_4FGLDR3}, each adding two years of data.
This new Data Release (DR4) covers 14 years of data.
This note focuses on what has changed since DR3, and describes the results. The reader is referred to the 4FGL and DR3 papers for the detailed methodology, and the official reference remains the DR3 paper.

Section~\ref{lat_and_background} describes the data and the updates to the diffuse model, Section~\ref{catalog_main} the updates to the analysis and the results, and Section~\ref{dr4_assocs} the updates to the associations.

\section{Instrument \& Background}
\label{lat_and_background}

\subsection{The LAT Data}
\label{LATData}

The data for the 4FGL DR4 catalog were taken during the period 4 August 2008 (15:43 UTC) to 2 August 2022 (21:53 UTC) covering 14 years.
During most of this time, \Fermi was operated in sky-scanning survey mode, with the viewing direction rocking north and south of the zenith on alternate orbits such that the entire sky is observed every $\sim$3 hours.
Since 8 April 2018 it is operated in partial sky-scanning mode\footnote{See \url{https://fermi.gsfc.nasa.gov/ssc/observations/types/post_anomaly/}.}. This has little impact on the integrated sky coverage relevant to the source catalog.

As in 4FGL, intervals around solar flares and bright $\gamma$-ray bursts (GRB) were excised. During the last two years, 54 ks were cut due to bright solar flares between July 2021 and January 2022, and 1 ks around 3 new bright GRBs.
The current version of the LAT data and Instrument Response Functions remains P8R3\_V3 \citep{LAT13_P8, LAT18_P305}.
The energy range remains 50~MeV to 1~TeV, and the data are split over the same 19 components with the same zenith angle selections as in DR3.

\subsection{Model for the Diffuse Gamma-Ray Background}
\label{DiffuseModel}

\begin{figure}[!ht]
   \centering
   \begin{tabular}{cc}
   \includegraphics[width=0.5\textwidth]{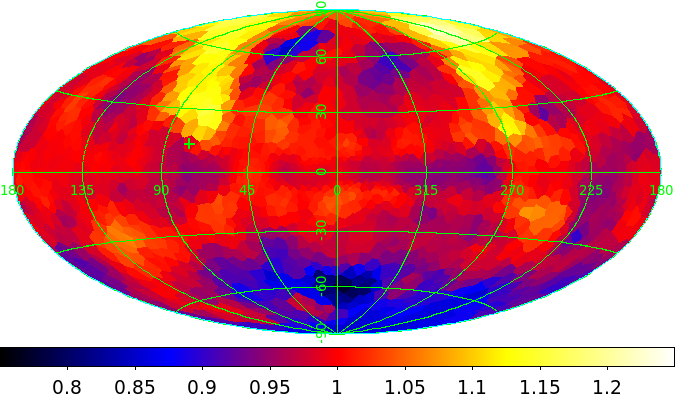} & 
   \includegraphics[width=0.5\textwidth]{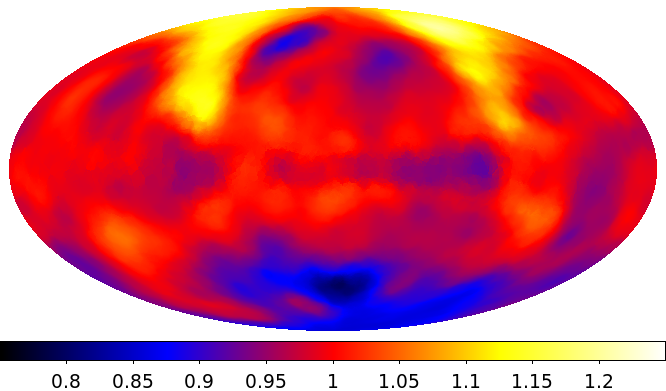}
   \end{tabular}
\caption{Normalization of the Galactic diffuse component in Hammer-Aitoff projection in Galactic coordinates. Left: Value obtained in each RoI over 12 years. Right: Smoothed version for modulating the interstellar emission model. The scale is the same on both sides.}
\label{fig:galnorm_interp}
\end{figure}

We used the same model for the interstellar emission (gll\_iem\_v07) and the same isotropic spectrum (iso\_P8R3\_SOURCE\_V3\_v1) as in DR3.
We realized however that we could improve on the procedure used up to DR3.
In order to avoid the sharp jumps between neighboring Regions of Interest (RoIs) with independent normalizing parameters, we rescaled the full model smoothly from the parameter mapping obtained at RoI level.
To that end we adopted the following procedure:
\begin{enumerate}
\item We chose to fix the normalization of the isotropic component to 1 meaning that this component is truly isotropic for this analysis. We instead modulated the Galactic component only. In order to leave the same level of freedom we applied a LogParabola\footnote{See \url{https://fermi.gsfc.nasa.gov/ssc/data/analysis/scitools/source_models.html\#LogParabola}. The normalization is called $N_0$ there.} (LP) modulation to each RoI (setting the reference energy $E_0$ to 1~GeV). After refitting all sources and iterating we obtained a set of ($K_i$, $\alpha_i$, $\beta_i$) triplets covering the entire sky, where $K_i$, $\alpha_i$ and $\beta_i$ are the normalization, index and curvature parameters of the LP model in RoI $i$. Their distributions are summarized in Table~\ref{tbl:LPMod}.
\item To go from RoIs to a full mapping of each point ($l$, $b$) in the sky, we defined weights of the surrounding RoIs $w_i = \max(D_i,R_i,2\degr)^{-2} \sigma_i^{-2}$, where $D_i$ is the distance between ($l$, $b$) and the center of RoI $i$, $R_i$ is the RoI core radius \citep[see item 5 in \S 3.2 of][]{LAT20_4FGL} and $\sigma_i$ is the precision on the normalization estimate in the RoI. The $\sigma_i^{-2}$ term is the usual statistical weighting. The reasoning behind the first term is to set a weight that decreases with distance from the RoI (the $D_i$ term), but remains constant while inside the RoI core (the $R_i$ term) and does not increase too much near small RoIs (the $2\degr$ term).
\item The neighboring RoIs were ranked by increasing distance to the current point. We derived a smooth map of the normalization parameter $K(l,b) = \Sigma_i w_i(l,b) K_i$. The sum was truncated when the precision on $K(l,b)$ reached the target value of $10^{-3}$ (a little below the best precision on individual RoIs, reached in the Galactic Ridge) or the number of RoIs reached 15 (so that the smoothing remained reasonably local). The result is shown in Fig.~\ref{fig:galnorm_interp}.
\item We obtained the smooth maps of $\alpha$ and $\beta$ using the same weights (and capping the sum at the same number of RoIs) as for the normalization, at each position in the sky.
\item We modulated the original Galactic model by $A(l,b,E) = K(l,b) \left (\frac{E}{E_0}\right )^{-\alpha(l,b) - \beta(l,b) \ln(E/E_0)}$ over the sky and at all energies $E$.
\end{enumerate}

\begin{deluxetable*}{lrrrr}
\tabletypesize{\scriptsize}
\tablecaption{LP Modulation of the Interstellar Emission model
\label{tbl:LPMod}}
\tablewidth{0pt}
\tablehead{
\colhead{}&
\colhead{$K$ (normalization at 1~GeV)}&
\colhead{$\alpha$ (index)}&
\colhead{$\beta$ (curvature)}&
}

\startdata
Mean value & 0.986 & 0.003 & 0.005 \\
Standard deviation & 0.055 & 0.019 & 0.007 \\
\enddata

\tablecomments{~Mean and standard deviation over the entire sky of the three parameters of the LP modulation of the interstellar emission model. The normalization has a North/South asymmetry (Figure~\ref{fig:galnorm_interp}). The index tends to be negative (harder model) in the Galactic plane and positive at high latitudes. The curvature tends to be positive (downwards) everywhere. The overall amplitude of the modulation is modest.}

\end{deluxetable*}

This rescaled model was derived from the DR3 data set, list of seeds and RoIs before we started working on DR4. Since there is no reason why it should change with time we applied it directly to the DR4 data set, list of seeds (\S~\ref{catalog_detection}) and reoptimized RoIs (\S~\ref{catalog_significance}). In the maximum likelihood fitting the isotropic component was fixed, but we kept the same power-law modulation of the rescaled Galactic model as before, with the same prior widths (0.03 and 0.02) but prior means set to 1 and 0 for the normalization and the index, respectively. We quantified the effect by comparing the DR4 results with the results obtained starting from the same data set and list of seeds, but fitting the Galactic and isotropic diffuse components as in DR3. The resulting parameter distributions are much narrower than before (scatters of 0.015 and 0.008 instead of 0.051 and 0.020 on the normalization and the index, respectively) but they remain broader than expected from pure statistical fluctuations (0.004). We also checked that the log(Likelihood) improved on average.

The effect on sources was measurable, and mostly due to the LP modulation of the Galactic diffuse and fixing the isotropic component, rather than to the smoothing. The source significance increased by $0.02 \sigma$ on average, leading to 71 more sources above threshold. The number of curved spectra decreased by 245. The energy flux increased by about 2\%, with a scatter around 10\% ($0.5 \sigma$).

The LP modulation of the original diffuse model is anchored in the 0.1 to 10~GeV range, where most of the events are. This implies that its behavior above 10~GeV is largely an extrapolation of the low-energy range and does not accurately reflect the high-energy range. This has no impact on point sources, which depend little on the diffuse model above 10~GeV because the radius of the Point Spread Function (PSF) is narrower than $0.15\degr$ (68\% containment). But it impacts extended sources. The largest effect is observed in FGES J1036.3$-$5833 in Carina, which is a large ($4.9\degr$ diameter) and hard (\texttt{LP\_EPeak} = 25~GeV) extended source on top of a +50\% modulation of the Galactic diffuse at 100~GeV. Its energy flux decreased by nearly 10\% between DR3 and DR4.

The model for the emission of the Sun and the Moon was kept the same, neglecting the modulation of the emission along the solar cycle. It was only updated to account for the additional two years of data.

\section{Construction of the Catalog}
\label{catalog_main}

Most of the steps were identical to DR3.
We use the same Test Statistic TS = 2 $\log (\mathcal{L} / \mathcal{L}_0)$ to quantify how significantly a source emerges from the background, comparing the maximum value of the likelihood function $\mathcal{L}$ including the source in the model with $\mathcal{L}_0$, the value without the source.

\subsection{Extended sources}

\begin{deluxetable*}{llllcl}
\tabletypesize{\scriptsize}
\tablecaption{Extended Sources Entered or Modified in the 4FGL-DR4 Analysis
\label{tbl:extended}}
\tablewidth{0pt}
\tablehead{
\colhead{DR4 Name}&
\colhead{Extended Source}&
\colhead{Origin}&
\colhead{Spatial Form}&
\colhead{Extent [deg]}&
\colhead{Reference}
}

\startdata
J0205.6+6449e & 3C 58 & New & Gaussian & 0.045 & \citet{3C58_Li18} \\
J0822.1$-$4253e & Puppis A & 3FHL & Map & 0.45 & \citet{PuppisA_Mayer22} \\
J1119.0$-$6127e & SNR G292.2$-$0.5 & New & Gaussian & 0.15 & \citet{HGPS_2018} \\
J1925.2+1618e & SNR G51.3+0.1 & New & Gaussian & 0.22 & \citet{G51.3_Araya21} \\
J1954.4+3252e & CTB 80 & New & Disk & 0.65 & \citet{CTB80_Araya21} \\
J2051.0+3049e & Cygnus Loop & 2FGL & Map & 1.65 & \citet{CygnusLoop_Tutone21} \\
\enddata

\tablecomments{~List of updated and new sources that have been modeled as spatially extended. The Origin column gives the name of the \Fermilat catalog in which that extended source was first introduced (with a different template). The Extent column indicates the radius for Disk (flat disk) sources, the 68\% containment radius for Gaussian sources, and an approximate radius for Map (external template) sources.}

\end{deluxetable*}

We updated the extended sources (Table~\ref{tbl:extended}). Multiwavelength templates were adopted for the Cygnus Loop and Puppis A, reducing the residuals considerably with respect to the previous geometric models while eliminating the two point sources inside the Cygnus Loop, which were only excess emission on top of the previous ring template. For the Cygnus Loop, we used only the single UV template rather than the more complex two-component representation advocated by \citet{CygnusLoop_Tutone21}. For Puppis A, we chose the eROSITA image above 1 keV \citep{PuppisA_Mayer22} because above that energy the bias on the X-ray intensity due to interstellar absorption is minimal.

In conjunction with the preparation of the Third \Fermi-LAT Pulsar Catalog \citep[3PC,][]{LAT23_3PC}, we added three extended sources around known pulsars, when their spectral analysis indicated a high-energy excess. 3C 58 is certainly a pulsar wind nebula (PWN) but the $\gamma$-ray emission in CTB 80 and G292.2$-$0.5 could also come partly from a surrounding supernova remnant (SNR). The broadest one, CTB 80 next to PSR J1952+3252, was represented by the disk fit to the LAT data by \citet{CTB80_Araya21}, eliminating 4FGL J1955.1+3321. The two smallest were represented by Gaussian shapes. For G292.2$-$0.5 (next to PSR J1119$-$6127) we used the TeV size from the H.E.S.S. Galactic Plane Survey. For 3C 58 (next to PSR J0205+6449) we used the radio size \citep{3C58_radio06}. For simplicity we ignored the elongation (the source is smaller than the PSF, so it is not critical). 3C 58 is so small (Gaussian $\sigma = 0.03\degr$) that the pulsar and the PWN cannot be separated spatially when building the spectral energy distribution (SED). The spectral indices of the two sources in each band (derived from the overall spectra) differ, so a best fit formally exists, but large fluctuations remain in the individual spectra.

Finally, we added the SNR G51.3+0.1 as detected by \citet{G51.3_Araya21}, deleting the three point sources that filled this area in DR3.

\subsection{Transient sources}
\label{catalog_transients}

Since many $\gamma$-ray sources are variable, a number of transients, detectable on short time scales, are diluted over many years and do not appear in the general source catalogs. GRBs are so short that they can be excised by simple time selection, but the other transients remain in the data. If they do not appear significant over the full time range, they do not contaminate very much neighboring sources on average, but they can possibly contaminate the light curves (i.e., they can be significant over one particular year).

\begin{deluxetable*}{lllclcl}
\tabletypesize{\scriptsize}
\tablecaption{Transient Sources Entered in the 4FGL-DR4 Analysis
\label{tbl:transients}}
\tablewidth{0pt}
\tablehead{
\colhead{DR4 Name}&
\colhead{Original name}&
\colhead{Flare date}&
\colhead{Class}&
\colhead{Localization}&
\colhead{Index}&
\colhead{Reference}
}

\startdata
J0105.5+1912 & ASV 33 & 09/2020 & bcu & ASV & 2.3 & \nodata \\
J0351.6+2921 & ASV 52 & 01/2022 & bcu & ASV & 2.2 & \nodata \\
J0358.4$-$5446 & YZ Ret & 07/2020 & NOV & Optical & 2.5 & \citet{YZRet_Sokolovsky22}\\
J1117.5$-$4839 & 1FLT J1117$-$4839 & 10/2009 & bcu & 1FLT & 2.4 & \citet{2021_1FLT} \\
J1146.4$-$0926 & 1FLT J1146$-$0926 & 02/2017 & bcu & 1FLT & 2.5 & \citet{2021_1FLT} \\
J1513.6$-$2830 & 1FLT J1513$-$2830 & 06/2014 & \nodata\tablenotemark{a} & 1FLT & 2.5 & \citet{2021_1FLT} \\
J1533.7$-$2130 & 1FLT J1533$-$2130 & 10/2016 & \nodata & 1FLT & 2.2 & \citet{2021_1FLT} \\
J1626.0+5436 & 1FLT J1626+5436 & 10/2015 & bcu & 1FLT & 2.8 & \citet{2021_1FLT} \\
J1820.8$-$2822 & V5856 Sgr & 11/2016 & NOV & Optical & 2.3 & \citet{V5856Sgr_Li17} \\
J1937.1$-$5509 & 1FLT J1937$-$5509 & 11/2016 & \nodata\tablenotemark{a} & 1FLT & 2.6 & \citet{2021_1FLT} \\
J2010.2$-$2523 & 1FLT J2010$-$2523 & 09/2014 & fsrq & 1FLT & 2.8 & \citet{2021_1FLT} \\
J2023.5+2046 & V339 Del & 08/2013 & NOV & Optical & 2.2 & \citet{Novae_2014} \\
J2102.1+4546 & V407 Cyg & 03/2010 & NOV & Optical & 2.4 & \citet{V407Cyg_2010} \\
J2240.9$-$1825 & ASV 24 & 05/2022 & bcu & ASV & 2.3 & \nodata \\
\enddata

\tablecomments{~List of transient sources, undetectable over 14 years, which have been entered into DR4. The ``Flare date'' column indicates the month and year of the flare. The Class column refers to the acronyms defined in Table 7 of \citet{LAT20_4FGL}. The Localization column indicates the origin of the localization. The Index column indicates the photon index of the power-law spectral model.}
\tablenotetext{a}{The 1FLT catalog reports a possible blazar association (not found by our automatic procedure) based on visual examination of the contents of the error ellipse.}

\end{deluxetable*}

In order to mitigate this effect we introduced in the DR4 catalog the brightest transients, which get close to or above the TS $>$ 25 significance level over one year. We took them from three lists of transients:
\begin{itemize}
\item Novae are detected over weekly time scales and have a single flare, so they are strongly diluted over one year. Four (V1369 Cen 2013, V5668 Sgr 2015, V906 Car 2018 and RS Oph 2021) were bright enough to be in DR4 or a previous data release already. V1369 Cen (4FGL J1353.3$-$5910) and V5668 Sgr (4FGL J1837.6$-$2904) are possibly contaminated by another source, because their light curves show a faint but non-zero signal besides the peak at the time of the nova. Among the other LAT-detected novae, we found that only those detected at TS $>$ 200 remain significant over one year, leaving only four of them.
\item \citet{2021_1FLT} looked systematically for transients outside the Galactic plane and not already in 4FGL DR2 on a monthly timescale. The resulting 1FLT catalog contains 142 such transients, mostly associated with AGNs. This was later updated to 12 years (1FLTi) and DR3. A number of them have entered DR3 or DR4 naturally (because they became brighter in recent years). Among those that did not, we found that those that are significant over one year are detected at least twice, and at a maximum TS of at least 60. This criterion resulted in seven entries.
\item Because the 1FLT catalog was not yet updated to 14 years, we used another  method based on the assessment of monthly variability (ASV for All-Sky Variability) for the last two years. The principle of this method is to first measure the monthly light curves of the 4FGL-DR3 variable sources and then to compare, for each monthly time bin, the observation to the prediction of a sky model containing only 4FGL-DR3 sources. The comparison is done by computing TS maps, which allow detecting flares of $\gamma$-ray sources not reported in 4FGL DR3. Although the ASV and 1FLT approaches differ, we have checked that the two methods detect the same flares above our threshold at TS = 60. We applied the same criterion, leading to three additional entries.
\end{itemize}

The results of that selection are given in Table~\ref{tbl:transients}. We entered those transients like the DR3 sources, so that they are preserved even at TS $<$ 25. Because they are very faint when averaged over 14 years, following the general rule and leaving their power-law indices to vary would have led to ill-defined values that would have then been propagated to all years, including that of the flare. Instead we fit them in the year during which they were strongest and fixed those indices in the fit over 14 years (and the other years) so that only the normalizations were allowed to vary. Therefore all the transients have no error on \texttt{PL\_Index}. The parameters of the curved spectral shapes were filled when the fits (over 14 years) converged.

We fixed the novae at their precise optical positions, so they do not have error ellipses. The error ellipses of the 1FLT and ASV transients, on the other hand, are the $\gamma$-ray localization during the brightest flare.

Three transients were not found formally variable by the variability index. Nova YZ Ret 2020 is very close ($< 0.3\degr$) to another source (4FGL J0358.5$-$5432) so it has large errors. The other two (4FGL J1513.6$-$2830 from 1FLT and J2240.9$-$1825 from ASV) have only moderately close neighbors but the contrast between the brightest year and the average is only a factor of 6, less than the other transients.

\subsection{Detection and Localization}
\label{catalog_detection}

The source detection followed the same approach as in 4FGL-DR3.
It used $pointlike$ and a specific diffuse model in which the non-template features are estimated differently from the 4FGL diffuse model.
It started from the 4FGL-DR3 sources, relocalized them over 14 years of data, looked for peaks in the residual TS maps generated for several spectral shapes, introduced those in the model, refit and iterated over the full procedure.
The result, called uw1410, contained more than 11,700 seed sources at TS $>$ 10.

We associated those seeds with the DR3 sources, keeping the DR3 original positions (consistent with their names). We eliminated seeds too close to a bright source and inside extended sources.
4FGL J2107.7+3529 (with a large error ellipse) was split into two uw1410 sources.
Seven faint DR3 sources were replaced by new nearby uw1410 sources that gathered most of the flux (leaving the original DR3 source with TS $<$ 9) when fit together.

Seven sources (6 AGNs and Cyg X-3) were much better localized in uw1410 than in DR3, because they became much more active after July 2020. We adopted the new error ellipses and set their DataRelease to 4.
During the investigation of the transients (\S~\ref{catalog_transients}) we noted that three flares close to existing sources were more compatible with the uw1410 than the DR3 positions, so we moved the sources to the uw1410 positions. One of them (4FGL J0427.3+0504, which becomes 4FGL J0426.5+0517) was relatively bright (TS $\sim$ 300) and gained an association at its new position (to PKS 0423+051).

The all-sky verification \citep[\S~3.6 of][]{LAT22_4FGLDR3} showed that two positive residuals had become very large \citep[PS $>$ 10, where the $p$-value statistic PS is defined in][]{PSmap} after following strictly the policy of deleting seeds inside extended sources. One in Carina corresponded to a uw1410 seed so we added it to the catalog as 4FGL J1038.8$-$5848, even though its spectrum (very similar to that of the underlying extended emission) and the morphology of the $\gamma$-ray emission indicate that it is probably not a point source. The second large residual, in Cygnus, did not correspond directly to a particular uw1410 seed, so we added a source as 4FGL J2033.0+3900 at the peak of the residual, without an error ellipse. We added another uw1410 seed inside FHES J2304.0+5406 corresponding to a PS = 8.7 residual. That one has the same hard spectrum as the extended source, but is very peaked and point-like. 

Seventeen pulsars known from timing analysis but too faint to be detected in uw1410 were entered at fixed positions. Two (PSR J1731$-$4744 and J1909$-$3744) survived the TS $>$ 25 cut and are included in DR4 at their timing positions with no error ellipse. In reverse, three pulsars known from timing were detected by uw1410 but did not reach the final TS $>$ 25 cut (PSR J0117+5914, J0154+1833 and J1946+3417).
In the end 4846 new seeds were entered to the $gtlike$ source characterization in addition to 6569 DR3 point sources and the 82 extended sources.

We reassessed the systematics on localization, and increased them a little. The absolute 95\% systematic error was set to $28.5\arcsec$ (instead of $25\arcsec$ and $27\arcsec$ at high and low latitudes, respectively, in DR3), while the systematic factor was set to 1.075 (instead of 1.06) at high latitude and 1.48 (instead of 1.37) close to the Galactic plane ($|b| < 5\degr$).
In the incremental approach those new numbers were applied only to the new sources (and those listed above for which we adopted the uw1410 localization).

\subsection{Spectral Shapes}
\label{catalog_spectra}

\begin{figure}[!ht]
   \centering
   \begin{tabular}{cc}
   \includegraphics[width=0.49\textwidth]{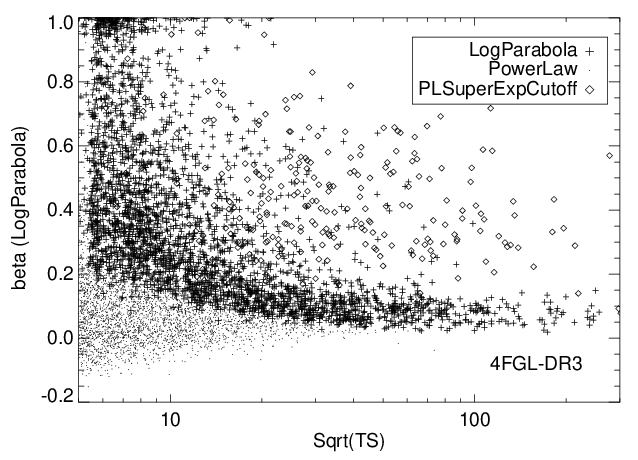} & 
   \includegraphics[width=0.49\textwidth]{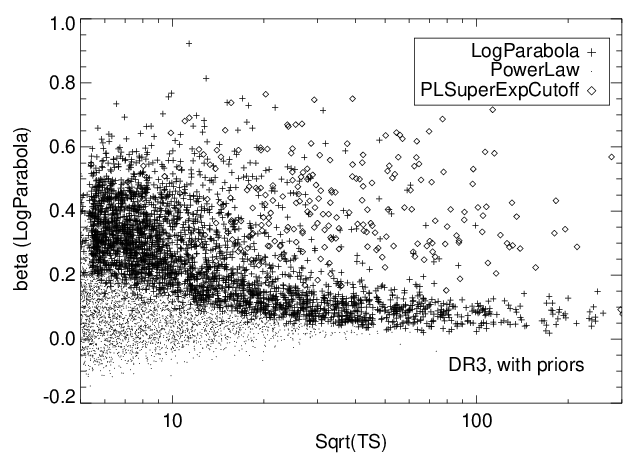}
   \end{tabular}
   \caption{Curvature parameter $\beta$ of the LP model plotted as a function of $\sqrt{\rm TS}$. The uncertainty increases to the left, approximately as $1/X$. The parameter value was limited to 1. Plus signs are sources whose best model was LP. Dots are sources in which curvature was not significant (modeled as a power law). Diamonds are pulsars (the LP model was fit to all sources, irrespective of their best model, so its parameters are always available). Left: In the DR3 catalog. Right: Same scatter plot with the priors on curvature, also over 12 years.}
\label{fig:TS_beta}
\end{figure}

Since we introduced curved spectra in the 2FGL catalog \citep{LAT12_2FGL}, we noticed that faint sources too often end up at $\beta$ = 1, the maximum allowed in the fit (Fig.~\ref{fig:TS_beta}, left). This never happens for bright sources (large TS). Bright AGNs (modeled as LP) have $\beta \sim 0.1$, and even pulsars (modeled in the Fermitools as PLSuperExpCutoff4 or PLEC4) never reach $\beta$ = 1 when they are bright. The conclusion was that those strongly curved spectra were not physical, but the result of the low-energy confusion due to the broad PSF. Many faint sources are modeled as power laws for lack of evidence that their spectra are curved, even though most likely they are, just like the bright sources. Those power laws tend to overestimate the real spectrum at low energy, so that the curved neighbors will compensate by appearing more curved than they really are.

In order to avoid this behavior, we introduced Gaussian priors on the curvature parameters in the DR4 catalog. We based the mean of those priors on the average for bright sources, and chose the width narrow enough to avoid the very large curvatures, but broader than the natural scatter observed on bright sources. For the $\beta$ parameter in LP, we set the mean to 0.1 and the width to 0.3. For the ExpfactorS parameter in PLEC4, we set the mean to 0.6 and the width to 0.6 (the same width as on $\beta$ but a larger mean, accounting for the fact that ExpfactorS corresponds to 2 $\beta$). We tested this on the DR3 data set (Fig.~\ref{fig:TS_beta}, right) and it worked fine, eliminating all strongly curved spectra while leaving bright sources (TS $>$ 1000) essentially unaltered.

\begin{figure}[!ht]
   \centering
   \begin{tabular}{cc}
   \includegraphics[width=0.49\textwidth]{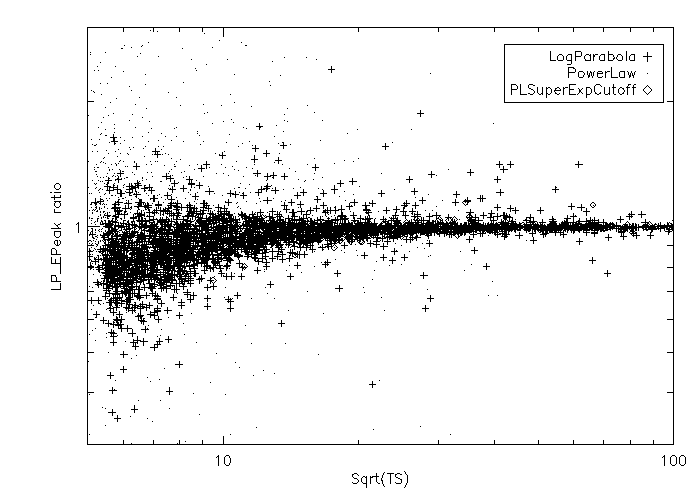} & 
   \includegraphics[width=0.49\textwidth]{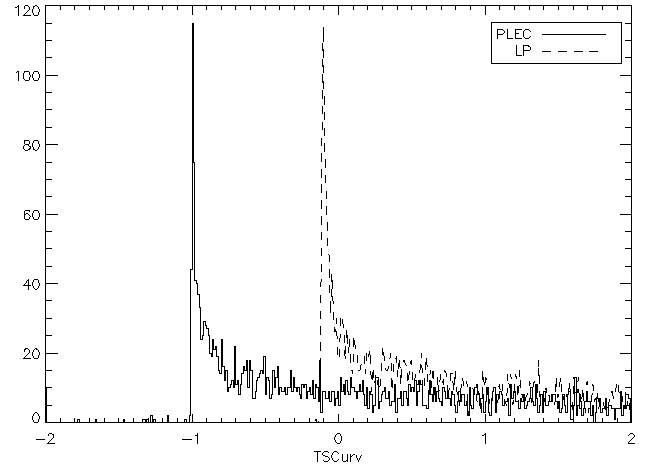}
   \end{tabular}
   \caption{Left: Ratio of the peak energy of the LP model with the priors on curvature to that in DR3, plotted as a function of $\sqrt{\rm TS}$. Plus signs are sources whose best model was LP. Dots are sources in which curvature was not significant (modeled as a power law). Diamonds are pulsars. Right: Distribution of TSCurv in the DR4 catalog, for the PLEC4 (peak at $-1$) and LP (peak at $-1/9$) spectral models. Convergence problems when attempting to fit the PLEC4 model caused 20 occurrences of PLEC\_TS$_{\rm curv} < -1$. Similar problems affected DR3 (30 PLEC\_TS$_{\rm curv} < 0$).}
\label{fig:TSCurv}
\end{figure}

We observed four minor side effects of that change:
\begin{itemize}
\item The number of curved spectra decreased by about 2\%, because the priors require a little more statistical evidence to reach the TS$_{\rm curv} > 4$ threshold. An even smaller ($<$ 1\%) fraction of sources (which were originally strongly curved) did not reach the TS $>$ 25 limit with the priors and disappeared entirely from the catalog.
\item The energy fluxes of the faint sources that became power laws, or in which $\beta$ decreased a lot, increased accordingly (because the energy flux is dominated either by low or high energies, depending on the spectral index, at which a strongly curved spectrum is lower). This effect reached as much as 20\% in individual sources. Collectively, however, the total energy flux in all sources increased by less than 1\%.
\item The peak energies decreased somewhat (Fig.~\ref{fig:TSCurv}, left). Overall, LP\_EPeak decreased by only 2\%, but that rises to 7\% for faint sources (TS $<$ 100) and to 15\% for faint sources whose best spectral model is LP, which are nearly all at $\beta > 0.1$ (Fig.~\ref{fig:TS_beta}), above the prior mean. This is because the high-energy part of the spectrum is better constrained than the low-energy part (more confusion at low energy). The prior mean at 0.1 on $\beta$ forces the curvature to be lower, but this translates into moving the model spectrum up more at low energy than at high energy, implying a lower EPeak.
\item Because of the prior, a pure PL spectrum ($\beta$ = 0) now gets a negative LP\_TS$_{\rm curv}$ = $-$(PriorMean/PriorWidth)$^2$ = $-$1/9. This is illustrated in Fig.~\ref{fig:TSCurv} (right). The effect is even stronger for PLEC4, because the prior mean and width on ExpfactorS are equal, so it can reach down to $-$1. These all get Sig$_{\rm curv}$ = 0 in the catalog, but there are many more values at 0 than in DR3 (618 for LP and 1592 for PLEC4). This can occur even for bright sources.
\end{itemize}

There are 277 pulsars in DR4, up from 255 in DR3. Most are modeled as PLEC4. Eight are not significantly curved and modeled as power laws.
The same 34 bright sources (28 pulsars and 6 blazars) as in DR3 were modeled as PLEC4 with free superexponential index $b$.

Because the priors make it a little more difficult to reach the TS$_{\rm curv} > 4$ threshold, the number of LP spectral shapes decreased from 3131 in DR3 to 3076 in DR4 (among which 152 are new sources). The number of PLEC4 spectra increased from 258 to 276 (two are new sources), thanks to the identification of new pulsars. Overall, the fraction of sources with a curved spectral model decreased from 51\% in DR3 to 47\% in DR4.

\subsection{Thresholding, Spectral Energy Distributions and Light Curves}
\label{catalog_significance}

We used the Fermitools package v2.2.0.
The likelihood weights were recomputed over 14 years of data, resulting in slightly smaller weights throughout.
We reoptimized all RoIs, resulting in 1991 RoIs containing up to 10 sources in their core.
As explained in \S~\ref{DiffuseModel}, the isotropic level was fixed to 1 in each RoI in the main fit, and only the interstellar emission level and its spectral bias were left free.

All DR3 sources were entered in the analysis with their DR3 spectral model, except the newly identified pulsars.
As in the previous incremental catalogs, they were not deleted from the model even if they had TS $<$ 25 at the end of the iteration process.
The resulting catalog contains 7195 entries, among which 546 are new, and 320 are DR3 sources or transients at TS $<$ 25.

The SEDs were obtained in the same way as in DR3 and over the same 8 bands, except the isotropic level was fixed to 1 as in the main fit.

The variability analysis proceeded in the same way as in DR3, now applied to 14 years (so the variability index threshold at 99\% confidence is 27.69).
For this analysis we left the isotropic free, because the weights for each year (the same as in DR3) are much larger than over 14 years so low energies are more important than in the main fit, and it is better to leave a little more freedom.
Indeed we noticed that the average values (over the entire sky) of the isotropic and interstellar emission levels were both slightly above 1 (around 1.03 for the isotropic and 1.003 for the interstellar). This implies that the Galactic curvature (optimized over the full interval) underestimates slightly the diffuse emission at a few hundred MeV. We also noticed a modulation over time of the average isotropic level (at the level of $\pm$ 2\%), which appears to follow the solar cycle. We interpret this as a variation of the residual charged-particle background in the $\gamma$-ray data set (minimum around solar maximum when the flux of cosmic rays at Earth is minimum).

The number of significantly variable sources increased from 1695 to 1825 (among which 47 are new sources). The fraction of variable sources remained constant at 25\%. Besides the known variable pulsars, three young pulsars are considered formally variable, with variability index between 35 and 50. They are PSR J1732$-$5049, J2043+2740 and J2208+4056. Two others are just above threshold, as expected by chance for a collection of nearly 300 pulsars.

The all-sky verification showed that positive residuals inside extended sources and negative residuals were at the same level as in DR3 (after adding the three seeds mentioned in \S~\ref{catalog_detection}).

\subsection{Analysis Flags}
\label{catalog_analysis_flags}

\begin{deluxetable*}{crrrl}

\tablecaption{Comparison of the numbers of flagged sources between DR3 and DR4. DR4 sources inherited from DR3 (``DR3 in DR4'') are considered separately. A given source can be flagged for several reasons, so the sum of individual rows is more than the number of flagged sources in the last row.
\label{tab:flags}}
\tablehead{
\colhead{Flag\tablenotemark{a}} & \colhead{DR3} & \colhead{DR3 in DR4} & \colhead{New in DR4} & \colhead{Meaning}
}

\startdata
  1  & 268 & 313 & 33 & $TS < 25$ with other model or analysis \\
  2  & 443 & 592 &  0 & Moved beyond 95\% error ellipse \\
  3  & 491 & 474 & 59 & Flux changed with other model or analysis \\
  4  & 460 & 463 & 78 & Source/background ratio $<$ 10\% \\
  5  & 677 & 687 & 90 & Confused \\
  6  & 317 & 316 & 38 & Interstellar gas clump (c sources) \\
  9  & 168 & 111 & 19 & Localization flag from {\it pointlike} \\
 10  &  46 &  49 &  1 & Bad spectral fit quality \\
 12  & 184 &   0 &  0 & Highly curved spectrum \\
 13  & 181 & 308 & 13 & $TS < 25$ \\
 14  & 549 & 547 & 40 & Soft Galactic Unassociated \\
 All & 2153 & 2278 & 222 & Any flag (\texttt{Flags} $>$ 0) \\
\enddata
 
\tablenotetext{a}{In the FITS file the values are encoded as individual bits in the \texttt{Flags} column, with Flag $n$ having value $2^{(n-1)}$.}

\end{deluxetable*}

The flags are recalled in Table~\ref{tab:flags} (see the DR3 paper for the detailed definitions), together with the numbers of sources flagged for each reason and their evolution since DR3.
The effect of the underlying interstellar emission model (IEM) was estimated by launching the procedure described in \S~\ref{catalog_significance} a second time using the same seeds but the previous IEM (gll\_iem\_v06).

We passed all new DR4 sources through the visual screening for diffuse features (Flag 6). That procedure flagged 38 sources. Flag 6 was not changed for DR3 sources.

The number of sources flagged with Flag 2 (inconsistent localization between the old DR4 position and uw1410) increased greatly, as well as the number of sources at low TS (Flag 13). This indicates that the current localizations (uw1410) have moved somewhat from the historical ones (DR1 to DR3). This implies that the concept of an incremental source catalog reaches its limit after six years, thus the next catalog will probably be an entirely new one.
The fraction of bad localizations among DR3 sources at high latitude is 8\%, rising to 12\% close to the Galactic plane. This exceeds the 5\% expected from the 95\% confidence errors. The reason is probably confusion (a fraction of sources tend to split into several), which is much greater in the Galactic plane.

Overall the fraction of sources with any flag set increased from 23\% in DR1 to 26\% in DR2, 32\% in DR3 and 35\% in DR4. Among the new DR4 sources, 70\% are flagged in the Galactic plane ($|b| < 10\arcdeg$), but only 26\% above the Galactic plane.

\subsection{The 4FGL-DR4 Catalog}
\label{dr4_description}

The catalog is available online\footnote{See \url{https://fermi.gsfc.nasa.gov/ssc/data/access/lat/14yr_catalog/}.}, together with associated products.
It contains 7195 entries (7194 sources, since the Crab Nebula has two entries), among which 546 are new point sources.
The source designation remains \texttt{4FGL JHHMM.m+DDMM}.
The format is the same as DR3.
However the \texttt{Unc\_Counterpart} column now contains the radius of the counterpart (rather than its localization uncertainty) whenever the counterpart was treated as extended during the association procedure. $\gamma$-ray point sources associated to extended counterparts can be recognized because they all have \texttt{ASSOC\_PRO\_BAY} = 0.8.

The detection threshold outside the Galactic plane decreased but remains a little above $1 \times 10^{-12}$ erg cm$^{-2}$ s$^{-1}$ in the 100~MeV to 100~GeV band.
The new sources are collectively similar to the new DR3 sources.
There are 11 with TS $>$ 100. Among those, eight are associated with blazars, and two are the excesses that we added to geometric extended sources in Carina and Cygnus (see \S~\ref{catalog_detection}). The last one (4FGL J0730.5+6720) has a $WISE$ AGN-like counterpart just below the association threshold of 0.8.

The number of new sources (546) is less than the number of sources introduced in DR2 (723) and DR3 (890). The reason why more sources entered the DR3 catalog was probably the software developments introduced on that occasion (priors on the diffuse parameters, lower TS$_{\rm curv}$ threshold for the LP spectral shape) and the fact that we have added more iterations to the detection process.
The reason why we add fewer sources in DR4 than in DR2 is less obvious. Of course the additional exposure is less in relative terms (+17\% instead of +25\%). The software changes (smooth diffuse modulation and priors on curvature) may also play a role.

\section{Associations}
\label{dr4_assocs}

 Since the DR3 release, some changes in associations or classes have been made:

\begin{itemize}
\item Taking the spatial extension of  globular clusters into account has led to six new associations (four sources were previously unassociated and two were classified as UNKs).
\item \cite{Cor22} found that 4FGL J1702.7$-$5655 is a candidate redback binary system. \cite{Swi22} reported that 4FGL J1408.6$-$2917 is a black-widow candidate. These two sources are both classified as binaries.   A low-mass X-ray  binary was discovered in 4FGL J1120.0$-$2204 \citep{Swi22a};
\item Pulsations have been discovered for 18 sources: 17 millisecond pulsars (MSPs), 8 of which were  previously classified as candidates and one as a globular cluster (NGC 6652), and one PSR. The list of detected pulsars in DR4 differs from 3PC  in that 13 PSRs and 5 MSPs are missing, while one PSR and one MSP are not in 3PC. Moreover, 4FGL J1023.7+0038 and 4FGL J1846.4$-$0258  are spatially  coincident with two other 3PC pulsars (PSR J1023+0038 and PSR J1846$-$0258) but are classified as a LMB and a PWN, respectively, on the basis of their $\gamma$-ray SEDs inconsistent with the pulsar class.
\item 12 new MSP candidates and two young pulsar candidates have been associated using the updated West Virginia University\footnote{\url{http://astro.phys.wvu.edu/GalacticMSPs/GalacticMSPs.txt}} or ATNF\footnote{\url{http://www.atnf.csiro.au/research/pulsar/psrcat/}} catalogs; 
\item The nova YZ Ret \citep{Sok22} is now the counterpart of the new source  4FGL J0358.4$-$5446. It was previously associated with  4FGL J0358.5$-$5432, whose counterpart is now the blazar candidate CRATES J035838$-$543404;
\item Four extended sources  (HESS J1804$-$216, HESS J1809$-$193, W 41, and SNR G106.3+02.7) with debated classes as either PWN or SNR, or showing composite features,  have been promoted to a new class, SPP (in capital letter in the CLASS1 column of the FITS file). These sources were previously classified as SPP candidates\footnote{These are sources of unknown nature but spatially overlapping with known SNRs or PWNe and thus candidate members of these classes.} (referred to as ``spp" in the FITS file). The classes of three extended sources  of unknown nature and associated with HESS J1507$-$622, HESS J1745$-$303, and HESS J1808$-$204 have been set to ``UNK".
\item Fifteen SPPs have additional associations from the LR method. Among these sources, six were previousy classified as ``UNK". The counterpart names provided by the LR method have been added to the ASSOC2 column. When available, this information has also been added for sources classified as SNR or GLC and associated with extended counterparts.
\item Four AGNs have changed classes in accordance with  4LAC-DR3  \citep{LAT22_4LACDR3}. NGC 6454 is classified as a radio galaxy, while TXS 0159+085, RX J0134.4+2638, and NVSS J121915+365718 are now classified as BL Lac objects (BLL);
\item Seven blazars of unknown types (BCUs) have been reclassified as BLLs following \cite{Rap23,paiano19,Pai20};
\item TXS 1433+205 has been found to be a distant Fanaroff-Riley II radio galaxy \citep{Pal23};
\item Two associations are now  high-confidence ones, while they were either low-confidence (4FGL J0153.3$-$1845) or omitted (4FGL J2306.6+0940).  The most probable counterpart of 4FGL J1912.2$-$3636 was spurious and has been removed.
\item Three associations from the LR-method  concerning  two BCUs (4FGL J0438.7$-$3441 and  4FGL J1950.6$-$5547) and one UNK  (4FGL J1925.1+1707) have been found spurious and removed. 
\item 4FGL J1744.9$-$2905 is now associated with NVSS J174456$-$290614, instead of NVSS J174230$-$282908.
\item The counterpart positions of four sources (4FGL J0601.3$-$7238, 4FGL J1910.8$-$6001,  4FGL J0524.8$-$6938, and 4FGL J0940.3$-$7610) have been fixed and others (for 4FGL J0955.3$-$3949 and  4FGL J1311.7$-$3430) have been improved.
\end{itemize}
\noindent In addition, six TeV associations have been added following  new detections reported in TeVCat\footnote{\url{http://tevcat.uchicago.edu/}}.  

Concerning the new sources in DR4, the association procedure  has been performed  by means of the Bayesian and likelihood-ratio (LR) methods along the lines of earlier releases. Associations have been obtained for 237 of the 546 new  point sources\footnote{These sources can be selected by  requiring DataRelease=4 and no entry in ASSOC\_4FGL.}. The Bayesian and the LR methods yield 201 and 139 associations respectively,  103 being in common. The  association fraction of 43\% is roughly consistent with that expected for the low-TS sources making up the DR4 sample, based on earlier results.  

These associations comprise:
\begin{itemize}
\item three pulsars with detected pulsations (one young pulsar PSR J1224$-$6407 and two MSPs, PSR J0737$-$3039A and PSR J1909$-$3744) and one MSP candidate PSR J1008$-$46;
\item  the four novae listed in Table \ref{tbl:extended} plus the bright RS Oph;
\item  one PWN candidate (Kes 75);
\item  one binary star (V918 Sco);
\item  one star-forming region (Sh 2-148);
\item 11 SPP candidates;
\item one globular cluster (Pal 6);
\item 14 sources of unknown nature (UNK, $|b|<10\arcdeg$ sources solely associated with the LR method from large radio and X-ray surveys);
\item 191 blazars including 25 BLLs, 28 flat-spectrum radio quasars (FSRQs) and 138 BCUs. 
The fraction of BCUs, which was 35\% in the 4FGL initial catalog, has risen to 72\% for the new blazars presented in subsequent releases (DR2 to DR4), due to the lack of available spectroscopic data for these blazar candidates. The new blazars (summed over all classes)  show a photon index distribution markedly different from that seen  in DR1:  while the latter exhibited a broad maximum about 2.2, the new-blazar  distribution peaks around 2.5, indicating the dominance of FSRQ-like sources \cite[see ][]{LAT22_4LACDR3}. This behavior is likely related to variability.  FSRQs show stronger and more frequent flares than BLLs in the LAT energy band, enhancing their detectability with respect to BLLs as livetime accumulates.     
\item 6 radio galaxies (2MASX J03204016+2727485, GB6 J1226+6406, 3C 293, PKS 1603+00, NGC 6061, LEDA 58287), one compact steep spectrum radio source (4C +76.03), one Seyfert (PKS 0000$-$160).
\end{itemize}
Two extended sources (SNR G292.2-0.5 and CTB 80) have been classified as SPPs, one as a SNR (SNR G51.26+0.11) and one as a PWN (3C 58).

We provide low-probability ($0.1<P<0.8$) associations for 43 sources and associations with {\sl Planck} counterparts for 15 others.  Three TeV associations (PKS 1413+135, RS Oph, 1RXS J195815.6$-$301119) have been added based on TeVCat.  

The overall fraction of soft Galactic unassociated sources \citep[SGUs, see][]{LAT22_4FGLDR3} has remained about the same ($\simeq$ 17\%) as in the previous two releases. A difference is observed when considering the Galactic latitude distributions, where the narrow component peaking on the Galactic plane \citep[the ``spike", see Figure 18 in][]{LAT22_4FGLDR3} is somewhat suppressed in DR4. This may indicate that, if SGUs are related to some mismodeled excess in diffuse emission, the spike excess has (at least  partially) been absorbed by the already-reported sources.

\begin{acknowledgments}
The \textit{Fermi} LAT Collaboration acknowledges generous ongoing support
from a number of agencies and institutes that have supported both the
development and the operation of the LAT as well as scientific data analysis.
These include the National Aeronautics and Space Administration and the
Department of Energy in the United States, the Commissariat \`a l'Energie Atomique
and the Centre National de la Recherche Scientifique / Institut National de Physique
Nucl\'eaire et de Physique des Particules in France, the Agenzia Spaziale Italiana
and the Istituto Nazionale di Fisica Nucleare in Italy, the Ministry of Education,
Culture, Sports, Science and Technology (MEXT), High Energy Accelerator Research
Organization (KEK) and Japan Aerospace Exploration Agency (JAXA) in Japan, and
the K.~A.~Wallenberg Foundation, the Swedish Research Council and the
Swedish National Space Board in Sweden.
 
Additional support for science analysis during the operations phase is gratefully
acknowledged from the Istituto Nazionale di Astrofisica in Italy and the Centre
National d'\'Etudes Spatiales in France. This work performed in part under DOE
Contract DE-AC02-76SF00515.

This work made extensive use of the ATNF pulsar  catalog\footnote{\url{http://www.atnf.csiro.au/research/pulsar/psrcat}}  \citep{ATNFcatalog}.  This research has made use of the \citet{NED1} which is operated by the Jet Propulsion Laboratory, California Institute of Technology, under contract with the National Aeronautics and Space Administration, of the SIMBAD database \citep{SIMBAD_2000} operated at CDS Strasbourg, France, and of archival data, software, and online services provided by the ASI Science Data Center (ASDC) operated by the Italian Space Agency.
We used the Manitoba SNR catalog \citep{Ferrand2012_SNRCat} to check recently published extended sources. We acknowledge the Einstein@Home project
for providing new pulsar associations through the dedicated efforts of the Einstein@Home volunteers. The Einstein@Home project is supported by the  NSF award 1816904.
\end{acknowledgments}

\software{Gardian \citep{Diffuse2}, GALPROP\footnote{\url{http://galprop.stanford.edu}} \citep{GALPROP17}, HEALPix\footnote{\url{http://healpix.jpl.nasa.gov/}} \citep{Gorski2005}, Aladin\footnote{\url{http://aladin.u-strasbg.fr/}}, TOPCAT\footnote{\url{http://www.star.bristol.ac.uk/\~mbt/topcat/}} \citep{Tay05}}

\facility{\Fermilat}

\bibliography{Bibtex_4FGL_v2}
 
\end{document}